\documentclass[12pt]{article}             
\begin{document}
\begin{center}
{\bf Deformed entropy and information relations for composite and noncomposite systems}\\
Vladimir N. Chernega,$^1$ Olga V. Man'ko$^1$ and Vladimir I. Man'ko$^{1,2}$\\
 $^1$P.N.~Lebedev Physical Institute\\
      Leninskii Prospect 53, Moscow 119991, Russia \\

$^2$Moscow Institute of Physics and Technology (State University)\\
Dolgoprudnyi, Moscow Region 141700, Russia\

      Email: omanko@sci.lebedev.ru
\end{center}
\begin{abstract}
The notion of conditional entropy is extended to noncomposite systems. The
$q$-deformed entropic inequalities, which usually are associated with
correlations of the subsystem degrees of freedom in bipartite systems, are
found for the noncomposite systems. New entropic inequalities for quantum
tomograms of qudit states including the single qudit states are obtained. The
Araki--Lieb inequality is found for systems without subsystems.
\end{abstract}
\vskip1cm

\noindent {\bf Key words:} marginal probability distribution, composite system,
entropy, deformation, conditional entropy, information relations\\
\noindent {\bf PACS}:
42.50.-p,03.65 Bz \vspace{0.4cm}

\section{Introduction}
The probability distributions are characterized by Shannon
entropy~\cite{Shanon}. The $q$-entropies~\cite{Reniy,Tsallis} containing an
extra parameter $q$ provide extra information on the probability
distributions. The state of quantum systems, identified with density matrices
\cite{Landau,vonNeuman} are characterized by von Neumann entropy. The
$q$-entropies also characterize the properties of quantum states. At complete
order in a classical system, the Shannon entropy is equal to zero. For
composite classical and quantum systems, there exist some inequalities related
to entropies of the system and its subsystems~\cite{Furuci,Furuci1,Petz-Vi}.

There exist the entropic and information inequalities, e.g., the subadditivity
condition, which is the inequality for von Neumann entropies of the
bipartite-system state and its two subsystem states~\cite{fromRusk}. For
three-partite systems, there exists the strong subadditivity condition, which
is the inequality for the von Neumann entropies of the composite system and
its subsystems~\cite{LiebRuskai}.  The nonnegativity of the Shannon mutual
information and quantum mutual information follows from the subadditivity
condition valid for composite systems.

Recently~\cite{RitaPS2014,TurinVova,TurinMA}, it was observed that all
entropic inequalities known for composite classical and quantum systems with
two or several random variables like, e.g., the subadditivity condition, can
be found also for the noncomposite system with only one random observable.

It is known (see, e.g., a recent review~\cite{NuovoCimento}) that the states
of quantum systems can be described in terms of fair probability
distributions, called quantum tomograms, which contain complete information on
the state density matrices.

The tomographic probability representation of quantum mechanics was suggested
in \cite{Mancini96}. The tomographic probability representation of classical
mechanics was suggested in \cite{OlgaJRLR1997}. Within the framework of this
representation, both classical and quantum states are described by the same
objects -- tomograms: the quantum states are determined by tomographic fair
probability distributions, the states of classical systems are determined by
classical tomograms. The analogous description of quantum spin states by the
probability distributions (spin tomograms) was suggested in
\cite{DodPLA,OlgaJETP}. In the tomographic probability representation, the
standard formulae for classical probability distributions can be easily
applied and compared with the corresponding quantum
ones~\cite{FoundFi,MamaMendesJRLR}.

The classical random variables are described within the framework of classical
probability theory \cite{Holevo}. The model of quantum mechanics based on the
classical probability distributions is elaborated in
\cite{Khrennikov,Khrennikov1,Khrennikov2}. Based on this fact and on the
tomographic-probability representation of quantum mechanics, one may use the
apparatus of classical probability theory to consider quantum correlations and
the entanglement phenomenon in quantum systems. For example, a specific map of
the classical probability distribution called the qubit portrait of qudit
states was introduced to study the entanglement phenomenon in
\cite{Vovf,Lupo}. The quantum correlations were studied within the framework
of tomographic probability representation of quantum
states~\cite{IbortPhys2007,maps}.

There exist different entropic inequalities for composite quantum systems
\cite{fromRusk,LiebRuskai,NielsonCH,PetzNielson,CorlenLieb2007,20pr,2014}.
Using the approach based on the portrait method it was observed in
\cite{RitaPS2014,OlgaVova,OlgaVova1} that the entropic inequalities valid for
composite systems can be extended to arbitrary systems including the systems
without subsystems. In \cite{CorlenLieb2007,20pr,2014}, some inequalities
associated with positive operators acting in the Hilbert space, which has the
structure of tensor product of Hilbert spaces, were studied.

In \cite{newineq}, new entropic inequalities for single qudit states were
obtained employing known properties of relative entropy of composite systems.
In \cite{subadditivity}, a new entropic inequality for states of the system of
$n\geq1$ qudits was derived, and a general statement on the existence of the
subbaditivity condition for an arbitrary probability distribution and an
arbitrary qudit-system tomogram was formulated. In \cite{Fedorov}, the
entropic inequalities and uncertainty relations for $q$-derformed entropy were
studied for noncomposite quantum systems realized by superconducting circuits
with the Josephson junction, and possible realizations of various quantum
logic gates of noncomposite quantum systems were discussed.

The aim of our work is to obtain new matrix inequalities for density matrices
of qudit states of noncomposite quantum systems which do not depend on the
tensor-product structure of the Hilbert space. The other goal of this paper is
to extend the notion of conditional entropy to the case of noncomposite
systems and to obtain the Araki--Lieb inequality for the single-qudit state,
as well as to obrain new inequalities for $q$-deformed entropy in the case of
noncomposite systems. Also we obtain a new chain relation for a single qudit
state.

This paper is organized as follows.

In the second section, we review the probability distributions and the
conditional entropies for one random variable. In the third section, we obtain
new entropic relations for qudit-state tomograms. In the fourth section we
find an analog of the Araki--Lieb inequality for an arbitrary density matrix
$\rho$ and consider an example of the matrix of a single-qudit state. In the
fifth section, we discuss the deformed subadditivity condition in the
classical and quantum cases and study quantum correlations expressed in terms
of the deformed information depending on global unitary transform. In
conclusion, we list our main results.

\section{The probability distributions and conditional entropies for one random variable}
The conditional entropy is the notion related to properties of the joint
probability distribution $P(j,k),$ $(j=1,2,\dots,n$, $k=1,2,\dots m)$ of two
random variables, where the first random variable describes degrees of freedom
of system~$A$ and second random variable describes degrees of freedom of
system~$B$. The joint probability distribution $P(j,k),$ determines the
conditional probability distributions, in view of Bayes rule,
\begin{equation}\label{eq.1}
P(j|k)=\frac{P(j,k)}{\sum_{j=1}^n P(j,k)}\,.
\end{equation}
The marginal probability distributions for the first and second random
variables read
\begin{equation}\label{eq.2}
{\cal P}_1(j)=\sum_{k=1}^m P(j,k),\quad {\cal P}_2(k)=\sum_{j=1}^n P(j,k).
\end{equation}
Shannon entropies associated to the probability distributions $P(j,k)$, ${\cal
P}_1(j)$, and ${\cal P}_2(k)$, as well as to the conditional probability
distribution $P(j|k)$ are
\begin{equation}\label{eq.3}
H(A,B)=-\sum_{j=1}^n\sum_{k=1}^m P(j,k)\ln P(j,k),\quad
H(A)=-\sum_{j=1}^n{\cal P}_1(j)\ln {\cal P}_1(j),\quad H(B)=-\sum_{k=1}^m{\cal
P}_2(k)\ln {\cal P}_2(k)
\end{equation}
and
\begin{equation}\label{eq.4}
H(A|k)=-\sum_{j=1}^n P(j|k)\ln P(j|k).
\end{equation}

The conditional entropy $H(A|B)$ is given by the average entropy
\begin{equation}\label{eq.5}
H(A|B)=\sum_{k=1}^m{\cal P}_2(k)H(A|k)=H(A,B)-H(B)
\end{equation}
Thus one has the equality
\begin{equation}\label{eq.6}
H(A,B)=H(A|B)+H(B),
\end{equation}
which is called the chain relation.

On the other hand, one can obtain an analogous relation for only one random
variable described by the probability distribution $P(s)$, $s=1,2,\dots N$,
where the integer $N=n m$. To show this possibility, following the approach
\cite{RitaPS2014}, we use the map of integers $1\leftrightarrow
11,\,2\leftrightarrow 12,\dots,m\leftrightarrow 1m, \,m+1\leftrightarrow
21,\,m+2\leftrightarrow 22,\dots,\,N-1\leftrightarrow nm-1,\,N\leftrightarrow
n m$; this means that the index $s$ in $P(s)$ is considered as double index $j
k$ where $j=1,2,\dots n$ and $k=1,2,\dots m$. Thus probability distribution
for one random variable is mapped onto the table $P(j,k)$ of nonnegative
numbers, which satisfies the normalisation condition
\begin{equation}\label{eq.8}
\sum_{s=1}^N P(s)=\sum_{j=1}^n\sum_{k=1}^m P(j,k)=1.
\end{equation}
Since all (\ref{eq.1})--(\ref{eq.8}) are formally the relations between the
$N=n m$ nonnegative numbers given by the table $P(j,k)$, the relations do not
depend on the interpretation of these numbers, say, as connected with a joint
probability distribution. They are valid also for the numbers $P(s)$
considered as the probabilities describing one randem variable but organized
as the table of numbers $P(j,k)$.

We give an example of $P(s)$ for four nonnegative numbers
$p_1,\,p_2,\,p_3,\,p_4$, such that $\sum_{s=1}^4 p_s=4$. One can introduce the
notation $P(1,1)\equiv p_1,\,P(1,2)\equiv p_2,\,P(2,1)\equiv p_3,\,
P(2,2)\equiv p_4$. Then one has analogs of all the probabilities given by
(\ref{eq.1}) and (\ref{eq.2}) as
\begin{equation}\label{eq.9}
{\cal P}_1(1)=p_1+p_2,\quad {\cal P}_1(2)=p_3+p_4,
\end{equation}
\begin{equation}\label{eq.10}
{\cal P}_2(1)=p_1+p_3,\quad {\cal P}_2(2)=p_3+p_4.
\end{equation}
Let us introduce two artificial subsystems $A$ and $B$ corresponding to
indices $j$ and $k$ in the table $P(j,k)$. Then we introduce the analogs of
conditional probability distributions. For example, all the numbers
\begin{equation}\label{eq.11}
P^A(1|1)=\frac{p_1}{p_1+p_3},\quad P^A(2|1)=\frac{p_3}{p_1+p_3}
\end{equation}
and
\begin{equation}\label{eq.13}
P^A(1|2)=\frac{p_2}{p_2+p_4},\quad P^A(2|2)=\frac{p_4}{p_2+p_4},
\end{equation}
can be considered as conditional probability distributions for subsystem~$A$.

These formulae provide the nonlinear maps of the probability four-vector $\vec
p=(p_1,p_2,p_3,p_4)$ onto two probability two-vectors, which are
\begin{eqnarray}
&&\vec p\rightarrow \vec P^A(1)=\frac{1}{p_1+p_3} \left(\begin{array}{c}
p_1\\
p_3
\end{array}\right),\qquad
\vec p\rightarrow \vec P^A(2)=\frac{1}{p_2+p_4} \left(\begin{array}{c}
p_2\\
p_4
\end{array}\right).\label{eq.15}
\end{eqnarray}
The Shannon entropies associated with the probability vectors~(\ref{eq.15})
read
\begin{eqnarray}
&&H^A(1)=-\frac{p_1}{p_1+p_3}\ln\frac{p_1}{p_1+p_3}-\frac{p_3}{p_1+p_3}\ln\frac{p_3}{p_1+p_3},\label{eq.16}\\
&&H^A(2)=-\frac{p_2}{p_2+p_4}\ln\frac{p_2}{p_2+p_4}-\frac{p_4}{p_2+p_4}\ln\frac{p_4}{p_2+p_4},\label{eq.16}
\end{eqnarray}
and the Shannon entropy associated with the four-vector $\vec p$ provides the
known chain relation for joint probability distribution, e.g., (\ref{eq.6}),
where we use the standard notation
\[H(A,B)=-p_1\ln p_1-p_2\ln p_2-p_3\ln p_3-p_4\ln p_4,\]
and the conditional entropy $H(A|B)$ reads
\begin{equation}\label{eq.18}
H(A|B) = (p_1+p_3)H^A(1)+(p_2+p_4)H^B(2);
\end{equation}
also
\begin{equation}\label{eq.19}
H(B)=-{\cal P}_2(1)\ln{\cal P}_2(1)-{\cal P}_2(2)\ln{\cal P}_2(2).
\end{equation}
The $q$-entropy of a bipartite system defined as
\begin{equation}\label{1A}
H_q(A,B)=-\sum_{j=1}^n\sum_{k=1}^m P^q(j,k)\frac{P^{1-q}(j,k)-1}{1-q}\,,
\end{equation}
for $q\to 1$, has the limit $H_1(A,B)=H(A,B)$.

The $q$-entropy $H_q(B)$ is defined as
\begin{equation}\label{2A}
H_q(B)=\frac{1}{q-1}\, \sum_{j=1}^m\left\{\left(\frac{P(j,k)}{\sum_{k'=1}^m
P(j,k')}\right)^q \left[\left(\frac{P(j,k)}{\sum_{k''=1}^m
P(j,k'')}\right)^{1-q}-1\right] \right\}.
\end{equation}
The conditional $q$-entropy $H_q(A|B)$ is the difference
\begin{equation}\label{3A}
H_q(A|B)=H_q(A,B)-H_q(B).
\end{equation}
Thus, we arrive at the chain relation~\cite{Furuci1,Petz-Vi,Rastegin}
\begin{equation}\label{4A}
H_q(A,B)=H_q(A|B)+H_q(B).
\end{equation}

One can write analogous relations using all permutations of numbers $p_s$.

From the consideration of the conditional entropies for the probability
distribution of one random variable $P(s)$ follows that the deformed chain
relation is valid for constructed `artificial' subsystems $A$ and $B$, e.g.,
described by the probability distributions given by (\ref{eq.9}) and
(\ref{eq.10}).

\section{Entropic relations for qudit tomograms}
Now we consider the qudit state tomograms. The tomograms are fair probability
distributions, which determine the density matrices of quantum states. In view
of this fact, the entropic relations for tomograms correspond to quantum
properties of qudits, in particular, to quantum correlations in multipartite
system states but also to quantum correlations in noncomposite system states.
The tomograms can be introduced for arbitrary Hermitian nonnegative matrix
$\rho$ with Tr$~\rho=1$.

The tomogram associated with the matrix $\rho$ reads
\begin{equation}\label{1.16}
w(s,u)=(u\rho u^+)_{ss}.
\end{equation}
Here $s=1,2,3,\dots,N=nm$ is an index characterising the basis in the linear
space where the density matrix is given. The tomogram is the standard
probability distribution depending on unitary matrix $u$ and it satisfies the
normalization condition
\begin{equation}\label{1.17}
\sum_{s=1}^N w(s,u)=1.
\end{equation}

One can introduce the deformed Shannon entropy for the tomogram, which is the
tomographic Tsallis entropy~\cite{Tsallis}
\begin{equation}\label{1.18}
H_q(u)=-\sum_{s=1}^N w(s,u)\frac{w^{q-1}(s,u)-1}{q-1}.
\end{equation}
For the joint probability distribution $P(j,k)$, there exists the deformed
subadditivity condition~\cite{Petz-Vi}, which we apply to the tomogram.

The deformed inequality which is a characteristics of the classical
probability $N$-vector $\vec w(u)$ with components $w(s,u)$ can be written in
the form $(q>1)$
\begin{equation}\label{1.19}
-\sum_{s=1}^N w(s,u)\frac{w^{q-1}(s,u)-1}{q-1}\leq-\sum_{j=1}^n
w_1(j,u)\frac{w_1^{q-1}(j,u)-1}{q-1}- \sum_{k=1}^m
w_2(k,u)\frac{w_2^{q-1}(k,u)-1}{q-1},
\end{equation}
where we have two probability vectors $\vec w_1(u)$ and $\vec w_2(u)$. The
components of these probability vectors are given as marginal probabilities
obtained from the table $P(j,k)$, where $j=1,2,\ldots,n$ and $k=1,2,\ldots,m$.
In this case, the table $P(j,k)$ is constructed from the
tomographic-probability distribution $w(s,u)$, $s=1,2,\ldots,n$, using the
same tool, which was used in the second section while considering the
probability distribution of one random variable $P(s)$ as the joint
probability distribution $P(j,k)$ of two artificial subsystems $A$ and $B$.
This means that instead of the probability distribution $P(s)$ we use the
tomographic-probability distribution $w(s,u)$, where the probabilities depend
on $N$$\times$$N$ unitary matrix $u$. The new quantum inequality~(\ref{1.19})
is valid for different systems.

We present the examples with two qubits and qudit with $j=3/2$.

The density matrix for two qubits is written in the basis $\mid
m_1m_2\rangle$, where $m_1,m_2=\pm 1/2$ in the Hilbert space $H=H_1\otimes
H_2$, which is the tensor product of two Hilbert spaces $H_1$ and $H_2$
corresponding to the states of the qubits. The matrix elements
$\rho_{m_1m_2,m'_1m'_2}$ provide the tomogram, which is the joint probability
distribution $w(m_1,m_2,u)$; thus, the index $s=1,2,3,4$ in the probability
vector $w(s,u)$ is mapped onto pairs
$1/2\,1/2,1/2\,-1/2,-1/2\,1/2,-1/2\,-1/2$.

Then the quantum inequality~(\ref{1.19}) for Tsallis $q$-entropy of the
two-qubit state reads
\begin{eqnarray}\label{1.19A}
-\sum_{m_1m_2=-1/2}^{1/2}w(m_1,m_2,u)\frac{w^{q-1}(m_1,m_2,u)-1}{q-1}\leq\nonumber\\
-\sum_{m_1=-1/2}^{1/2}w_1(m_1,u)\frac{w_1^{q-1}(m_1,u)-1}{q-1}
-\sum_{m_2=-1/2}^{1/2}w_2(m_2,u)\frac{w_2^{q-1}(m_2,u)-1}{q-1}\,,
\end{eqnarray}
where marginals $w_1(m_1,u)=-\sum_{m_2=-1/2}^{1/2}w(m_1,m_2,u)$ and
$w_2(m_2,u)=-\sum_{m_1=-1/2}^{1/2}w(m_1,m_2,u)$ are the tomograms for qubits,
if the unitary 4$\times$4-matrix $u$ is taken as the tensor product
$u=u_1\times u_2$ of local unitary transforms.

In the case of two qubits, one can get the chain rule for the entropies of two
subsystems given by equation~(\ref{eq.6}), where the entropy reads
\begin{equation}\label{1.20A}
H(A,B)=-\sum_{m_1m_2=-1/2}^{1/2}w(m_1,m_2,u)\ln w(m_1,m_2,u), \end{equation}
and the entropy for the second qubit $H(B)$ is
\begin{equation}\label{1.20B}
H(B)=-\sum_{m_2=-1/2}^{1/2}w_2(m_2,u)\ln w_2(m_2,u). \end{equation} The
conditional tomographic entropy $H(A|B)$ is given by equation~(\ref{eq.5}).

The second example under consideration is qudit with $j=3/2$; it provides the
same entropic inequalities, which are new for this system. We employ the map
of indices in the tomogram $w(s,u)$ interpreting index $s=1,2,3,4$ as the spin
projection $m=-3/2,-1/2,1/2,3/2$. This means that the tomogram $w(s,u)\equiv
w(m,u)$ satisfies the inequality
\begin{eqnarray}\label{1.19B}
-\sum_{m=-3/2}^{3/2}w(m,u)\frac{w^{q-1}(m,u)-1}{q-1}\leq\nonumber\\
-\sum_{j=1}^{2}\Omega_1(j,u)\frac{\Omega_1^{q-1}(j,u)-1}{q-1}
-\sum_{k=1}^{2}\Omega_2(k,u)\frac{\Omega_2^{q-1}(k,u)-1}{q-1},
\end{eqnarray}
where the probability distributions $\Omega_1(j,u)$ and $\Omega_2(k,u)$, with
$j.k=1,2$, are expressed in terms of the qudit tomograms according to
equations~(\ref{eq.9}) and (\ref{eq.10}) as
\begin{eqnarray}
\Omega_1(1,u)=w(-3/2,u)+w(-1/2,u),\qquad
\Omega_1(2,u)=w(1/2,u)+w(3/2,u),\label{1.21A}\\
\Omega_2(1,u)=w(-3/2,u)+w(1/2,u),\qquad
\Omega_2(2,u)=w(-1/2,u)+w(3/2,u),\label{1.21B}
\end{eqnarray}
Inequality~(\ref{1.19B}) is a new entropic inequality for the single qudit
state with $j=3/2$; it can be checked experimentally.

The realization of the qudit state can be provided either by the four-level
atomic state or by the Josephson-junction state in the quantum-circuit
experiments.

In the limit $q\to 1$, inequalities~(\ref{1.19A}) and (\ref{1.19B}) become the
subadditivity conditions for Shannon entropies determined by the tomograms.
Inequality~(\ref{1.19A}) provides the standard subadditivity condition for
bipartite system, and inequality~(\ref{1.19B}) determines the new
subadditivity condition for a single random variable.

Now we introduce the conditional entropy for the tomogram of the qudit state
with $j=3/2$. To do this, we write the $q$-entropy for the qudit state with
$j=3/2$ determined by the state tomogram as follows:
\begin{eqnarray}\label{1.19C}
H_{q}^{(3/2)}(u)&=&-\left\{w^{q}(3/2,u)\frac{w^{1-q}(3/2,u)-1}{1-q}+
w^{q}(1/2,u)\frac{w^{1-q}(1/2,u)-1}{1-q}\right.\nonumber\\
&&\left.+w^{q}(-1/2,u)\frac{w^{1-q}(-1/2,u)-1}{1-q}
+w^{q}(-3/2,u)\frac{w^{1-q}(-3/2,u)-1}{1-q}\right\}.
\end{eqnarray}
This expression for the $q$-entropy is equivalent to the left-hand side of
inequality (\ref{1.19B}).

The $q$-entropy related to the probability to obtain positive and negative
spin projections $\Omega_1(+,u)=w(1/2,u)+w(3/2,u)$ and
$\Omega_2(-,u)=w(-3/2,u)+w(-1/2,u)$ reads
\begin{equation}\label{1.21C}
H_q^B=-\Omega_1^{q}(+,u)\frac{\Omega_1^{q}(+,u)-1}{1-q}-
\Omega_1^{q}(-,u)\frac{\Omega_1^{q}(-,u)-1}{1-q}\,.
\end{equation}

Thus, we interpret an ``artificial'' subsystem $B$ as a set of events where
one has either only positive or only negative values of the spin projections
for the system with spin $j=3/2$ (qudit). The other ``artificial'' subsystem
$A$ is considered as a set of events where the modulus of the sum of the spin
projections is equal to unity. Quantum correlations of these two subsystems
correspond to the correlations of the different spin projections, which play
the role of different qubits in the qubit bipartite system.

We introduce the conditional entropy and the chain relation for the tomogram
of the qudit state with $j=3/2$ taking
\begin{eqnarray}\label{1.19D}
H_{q}(A|B)&=&\Omega_1^{q}(+,u)\frac{\Omega_1^{1-q}(+,u)-1}{1-q}+
\Omega_1^{q}(-,u)\frac{\Omega_1^{1-q}(-,u)-1}{1-q}\nonumber\\
&&-\left\{w^{q}(3/2,u)\frac{w^{1-q}(3/2,u)-1}{1-q}+
w^{q}(1/2,u)\frac{w^{1-q}(1/2,u)-1}{1-q}\right.\nonumber\\
&&\left.+w^{q}(-1/2,u)\frac{w^{1-q}(-1/2,u)-1}{1-q}
+w^{q}(-3/2,u)\frac{w^{1-q}(-3/2,u)-1}{1-q}\right\}.
\end{eqnarray}
Thus, one has the chain relation~(\ref{4A}) for the single qudit state with
$j=3/2$.

Analogous chain relations can be constructed for the other single qudits.

In the limit $q\to 1$, the chain relations become the entropic relations for
conditional tomographic Shannon entropies for the systems without subsystems.

\section{Araki--Lieb inequality for the single qudit state}

The subadditivity condition for the von Neumann entropy of the two-qudit state
with the density matrix $\rho(1,2)$ and the entropies for each qudit states
with the density matrices $\rho(1)=\mbox{Tr}_2\,\rho(1,2)$ and
$\rho(2)=\mbox{Tr}_1\,\rho(1,2)$, respectively, can be written in a form of
the matrix inequality~\cite{RitaPS2014}. In fact, the first qudit state with
$j=(n-1)/2$ and the second qudit state with $j=(m-1)/2$ are described by the
density matrix $\rho(1,2)$ of a block form
\begin{equation}\label{1}
\rho(1,2)=\left(\begin{array}{cccc}
R_{11}&R_{12}&\dots&R_{n1}\\
R_{21}&R_{22}&\dots&R_{2n}\\
\dots&\dots&\dots&\dots\\
R_{n1}&R_{n2}&\dots&R_{nn}
\end{array}\right),
\end{equation}
where blocks $R_{kl}$ ($k,l=1,2,\dots,n$) are $m\times m$-matrices and
$\rho(1,2)$ is the $N$$\times$$N$-matrix with $N=n m$. Then the density
$n$$\times$$n$-matrix of the first qudit state $\rho(1)$ reads
\begin{equation}\label{2}
\rho(1)=\left(\begin{array}{cccc}
\mbox{Tr}R_{11}&\mbox{Tr}R_{12}&\dots&\mbox{Tr}R_{n1}\\
\mbox{Tr}R_{21}&\mbox{Tr}R_{22}&\dots&\mbox{Tr}R_{2n}\\
\dots&\dots&\dots&\dots\\
\mbox{Tr}R_{n1}&\mbox{Tr}R_{n2}&\dots&\mbox{Tr}R_{nn}
\end{array}\right),
\end{equation}
and the density $m$$\times$$m$ matrix $\rho(2)$ is expressed in terms of
blocks $R_{kl}$ as
\begin{equation}\label{3}
\rho(2)=\sum_{k=1}^n R_{kk}.
\end{equation}
The subadditivity condition means that
\begin{eqnarray}
&&-\mbox{Tr}\left(\begin{array}{cccc}
R_{11}&R_{12}&\dots&R_{n1}\\
R_{21}&R_{22}&\dots&R_{2n}\\
\dots&\dots&\dots&\dots\\
R_{n1}&R_{n2}&\dots&R_{nn}
\end{array}\right)\ln\left(\begin{array}{cccc}
R_{11}&R_{12}&\dots&R_{n1}\\
R_{21}&R_{22}&\dots&R_{2n}\\
\dots&\dots&\dots&\dots\\
R_{n1}&R_{n2}&\dots&R_{nn}
\end{array}\right)\leq\nonumber\\
&&-\mbox{Tr}\left(\begin{array}{cccc}
\mbox{Tr}R_{11}&\mbox{Tr}R_{12}&\dots&\mbox{Tr}R_{n1}\\
\mbox{Tr}R_{21}&\mbox{Tr}R_{22}&\dots&\mbox{Tr}R_{2n}\\
\dots&\dots&\dots&\dots\\
\mbox{Tr}R_{n1}&\mbox{Tr}R_{n2}&\dots&\mbox{Tr}R_{nn}
\end{array}\right)\ln\left(\begin{array}{cccc}
\mbox{Tr}R_{11}&\mbox{Tr}R_{12}&\dots&\mbox{Tr}R_{n1}\\
\mbox{Tr}R_{21}&\mbox{Tr}R_{22}&\dots&\mbox{Tr}R_{2n}\\
\dots&\dots&\dots&\dots\\
\mbox{Tr}R_{n1}&\mbox{Tr}R_{n2}&\dots&\mbox{Tr}R_{nn}
\end{array}\right)\nonumber\\
&&-\mbox{Tr}\left(\sum_{k=1}^n R_{k k}\right)\ln\left(\sum_{k=1}^n R_{k k}\right).\label{4}
\end{eqnarray}
For the bipartite system state, the Araki--Lieb inequality provides a bound
for the difference of two subsystem quantum entropies; it reads
\begin{equation}\label{5}
-\mbox{Tr}\rho(1,2)\ln\rho(1,2)\geq|-\mbox{Tr}\rho(1)\ln\rho(1)+\mbox{Tr}\rho(2)\ln\rho(2)|.
\end{equation}
The inequality can be rewritten in the matrix form as follows:
\begin{eqnarray}
&&-\mbox{Tr}\left(\begin{array}{cccc}
R_{11}&R_{12}&\dots&R_{n1}\\
R_{21}&R_{22}&\dots&R_{2n}\\
\dots&\dots&\dots&\dots\\
R_{n1}&R_{n2}&\dots&R_{nn}
\end{array}\right)\ln\left(\begin{array}{cccc}
R_{11}&R_{12}&\dots&R_{n1}\\
R_{21}&R_{22}&\dots&R_{2n}\\
\dots&\dots&\dots&\dots\\
R_{n1}&R_{n2}&\dots&R_{nn}
\end{array}\right)\geq\nonumber\\
&&\left|-\mbox{Tr}\left(\begin{array}{cccc}
\mbox{Tr}R_{11}&\mbox{Tr}R_{12}&\dots&\mbox{Tr}R_{n1}\\
\mbox{Tr}R_{21}&\mbox{Tr}R_{22}&\dots&\mbox{Tr}R_{2n}\\
\dots&\dots&\dots&\dots\\
\mbox{Tr}R_{n1}&\mbox{Tr}R_{n2}&\dots&\mbox{Tr}R_{nn}
\end{array}\right)\ln\left(\begin{array}{cccc}
\mbox{Tr}R_{11}&\mbox{Tr}R_{12}&\dots&\mbox{Tr}R_{n1}\\
\mbox{Tr}R_{21}&\mbox{Tr}R_{22}&\dots&\mbox{Tr}R_{2n}\\
\dots&\dots&\dots&\dots\\
\mbox{Tr}R_{n1}&\mbox{Tr}R_{n2}&\dots&\mbox{Tr}R_{nn}
\end{array}\right)\right.\nonumber\\
&&\left.+\mbox{Tr}\left(\sum_{k=1}^n R_{k k}\right)\ln\left(\sum_{k=1}^n R_{k
k}\right)\right|.\label{6}
\end{eqnarray}
The Araki--Lieb entropic inequality written in the matrix form~(\ref{6}) is
valid for an arbitrary matrix $\rho$ given in the form (\ref{1}), which is a
nonnegative Hermitian matrix with the unit trace. In view of this fact, we
obtain an analog of the Araki--Lieb inequality for an arbitrary matrix $\rho$
of the form (\ref{1}), including the matrix of the single qudit state.

For example, for a qutrit state (or the spin state with $j=1$) with the
density matrix
\begin{equation}\label{7}
\rho=\left(\begin{array}{ccc}
\rho_{11}&\rho_{10}&\rho_{1-1}\\
\rho_{01}&\rho_{00}&\rho_{0-1}\\
\rho_{-11}&\rho_{-10}&\rho_{-1-1}
\end{array}\right),
\end{equation}
the subadditivity condition reads~\cite{OlgaVova}
\begin{eqnarray}
&&-\mbox{Tr}\left(\begin{array}{ccc}
\rho_{11}&\rho_{10}&\rho_{1-1}\\
\rho_{01}&\rho_{00}&\rho_{0-1}\\
\rho_{-11}&\rho_{-10}&\rho_{-1-1}
\end{array}\right)\ln\left(\begin{array}{ccc}
\rho_{11}&\rho_{10}&\rho_{1-1}\\
\rho_{01}&\rho_{00}&\rho_{0-1}\\
\rho_{-11}&\rho_{-10}&\rho_{-1-1}
\end{array}\right)\leq\nonumber\\
&&-\mbox{Tr}\left(\begin{array}{cc}
\rho_{11}+\rho_{-1 1}&\rho_{1 0}\\
\rho_{01}&\rho_{00}\end{array}\right)\ln\left(\begin{array}{cc}
\rho_{11}+\rho_{-1 1}&\rho_{1 0}\\
\rho_{01}&\rho_{00}\end{array}\right)\nonumber\\
&&-\mbox{Tr}\left(\begin{array}{cc}
\rho_{11}&\rho_{-1 1}\\
\rho_{-11}&\rho_{-1-1}\end{array}\right)\ln\left(\begin{array}{cc}
\rho_{11}&\rho_{-1 1}\\
\rho_{-11}&\rho_{-1-1}\end{array}\right).\label{8}
\end{eqnarray}
But for the qutrit state, one has the Araki--Lieb inequality
\begin{eqnarray}
&&-\mbox{Tr}\left(\begin{array}{ccc}
\rho_{11}&\rho_{10}&\rho_{1-1}\\
\rho_{01}&\rho_{00}&\rho_{0-1}\\
\rho_{-11}&\rho_{-10}&\rho_{-1-1}
\end{array}\right)\ln\left(\begin{array}{ccc}
\rho_{11}&\rho_{10}&\rho_{1-1}\\
\rho_{01}&\rho_{00}&\rho_{0-1}\\
\rho_{-11}&\rho_{-10}&\rho_{-1-1}
\end{array}\right)\geq\nonumber\\
&&\left|-\mbox{Tr}\left(\begin{array}{cc}
\rho_{11}+\rho_{-1 1}&\rho_{1 0}\\
\rho_{01}&\rho_{00}\end{array}\right)\ln\left(\begin{array}{cc}
\rho_{11}+\rho_{-1 1}&\rho_{1 0}\\
\rho_{01}&\rho_{00}\end{array}\right)\right.\nonumber\\
&&\left.+\mbox{Tr}\left(\begin{array}{cc}
\rho_{11}&\rho_{-1 1}\\
\rho_{-11}&\rho_{-1-1}\end{array}\right)\ln\left(\begin{array}{cc}
\rho_{11}&\rho_{-1 1}\\
\rho_{-11}&\rho_{-1-1}\end{array}\right)\right|.\label{9}
\end{eqnarray}
A qutrit is a system without subsystems. The Araki--Lieb inequality was known
for a system with two subsystems. Thus, we obtained a new entropic inequality
of a form of the Araki--Lieb inequality, which can be checked in the
experiments where the density matrix of qutrit is measured.

\section{Deformed subadditivity condition classical and quantum}
For any density matrix $\rho$, the quantum deformed entropy $S_q(\rho)$ reads
(see, e.g. \cite{Petz-Vi})
\begin{equation}\label{1.1}
S_q(\rho) =-\mbox{Tr}\rho\left(\frac{\rho^{q-1}-1}{q-1}\right).
\end{equation}
In the limit $q\rightarrow1$, the deformed entropy is equal to the von Neumann
entropy
\begin{equation}\label{1.2}
\lim_{q\rightarrow1}S_q(\rho)=-\mbox{Tr}\rho\ln\rho.
\end{equation}
For a bipartite system with subsystems 1 and 2 and the density matrix
$\rho(1,2)$, one has the inequality, which is deformed subadditivity
condition; it reads
\begin{equation}\label{1.3}
-\mbox{Tr}\,\rho(1,2)\,\frac{\rho^{q-1}(1,2)-1}{q-1}\leq
-\mbox{Tr}\,\rho(1)\,\frac{\rho^{q-1}(1)-1}{q-1}
-\mbox{Tr}\,\rho(2)\,\frac{\rho^{q-1}(2)-1}{q-1},
\end{equation}
where the density matrices of the subsystem states $\rho(1)$ and $\rho(2)$ are
\begin{equation}\label{1.4}
\rho(1)=\mbox{Tr}_2\rho(1,2),\qquad \rho(2)=\mbox{Tr}_1\rho(1,2).
\end{equation}
For $q\rightarrow1$, inequality~(\ref{1.3}) is the subadditivity conditions
for the von Neumann entropy
\begin{equation}\label{1.5}
-\mbox{Tr}\,\rho(1,2)\ln\rho(1,2)\leq-\mbox{Tr}\,\rho(1)\ln\rho(1)-\mbox{Tr}\,\rho(2)\ln\rho(2).
\end{equation}
For any nonnegative $N$$\times$$N$-matrix $\rho$ with $N=n m$ given in block
form~(\ref{1}), where $R_{j k}$ are $m$$\times$$m$-matrices, one has the
inequality
\begin{equation}\label{1.7}
-\mbox{Tr}\,\rho\,\frac{\rho^{q-1}-1}{q-1}\leq
-\mbox{Tr}\,R_1\,\frac{R_1^{q-1}-1}{q-1}
-\mbox{Tr}\,R_2\,\frac{R_2^{q-1}-1}{q-1}\,,
\end{equation}
where the $n$$\times$$n$-matrix $R_1$ has the form~(\ref{2}), and the
$m$$\times$$m$-matrix $R_2$ has the form~(\ref{3}).

Thus one has an analog of the deformed subadditivity condition for the single
qudit state with $j=(N-1)/2$. The $N$$\times$$N$-matrix $\rho$ can be
considered as a part of the $\tilde N$$\times$$\tilde N$-matrix $\tilde\rho$,
if one uses the appropriate number of extra zero columns and rows, i.e.,
$~\tilde\rho=\left(\begin{array}{cc} \rho&0\\0&0\end{array}\right).$ In view
of this, one has the general inequality for the matrix elements of the matrix
$\rho$ considering different product forms of the integer $\tilde N=\tilde
n\tilde m$. This means that we can derive several different entropic
inequalities starting from the $\tilde N$$\times$$\tilde N$-matrix
$\tilde\rho$ and considering different matrices $\tilde R_1$ and $\tilde R_2$
based on the block form of the matrix $\tilde\rho$.

The density matrix $\rho$ can be transformed, using unitary matrix $u$, to
become
\begin{equation}\label{1.10}
\rho\rightarrow\rho_u=u\rho u^+.
\end{equation}
The matrix $\rho_u$ which has the form (\ref{1}) also satisfies the
subadditivity condition~(\ref{1.7}), where the matrices $R_1$ and $R_2$ are
replaced by the matrices $R_1(u)$ and  $R_2(u)$. The matrices $R_1(u)$ and
$R_2(u)$ are given by formulae~(\ref{2}) and (\ref{3}), where the blocks $R_{j
k}$ are replaced by the blocks $R_{j k}(u)$ obtained from the matrix $\rho_u$.
Then one has a transformed inequality~(\ref{1.7})
\begin{equation}\label{1.11}
-\mbox{Tr}\,\rho\,\frac{\rho^{q-1}-1}{q-1}\leq
-\mbox{Tr}\,R_1(u)\,\frac{R_1^{q-1}(u)-1}{q-1}
-\mbox{Tr}\,R_2(u)\,\frac{R_2^{q-1}(u)-1}{q-1}\,,
\end{equation}
where the left-hand side of the inequality contains the matrix $\rho$, but in
right-hand side the matrices $R_1(u)$ and $R_2(u)$ depend on the unitary
matrix $u$.

In the limit $q=1$, one has the inequality
\begin{equation}\label{1.12}
-\mbox{Tr}\,\rho\ln\rho\leq-\mbox{Tr}\,R_1(u)\ln R_1(u)-\mbox{Tr}\,R_2(u)\ln
R_2(u).
\end{equation}
This inequality is valid for an arbitrary unitary matrix $u$. One can
introduce the quantum information, which depends on global unitary transform
\begin{equation}\label{1.13}
I(u)=\mbox{Tr}\,\rho\ln\rho-\mbox{Tr}\,R_1(u)\ln R_1(u)-\mbox{Tr}\,R_2(u)\ln
R_2(u)\geq0.
\end{equation}
The minimum value of the sum of entropies
\begin{equation}\label{1.14}
\Sigma(u_0)=-\mbox{Tr}\,R_1(u_0)\ln R_1(u_0)-\mbox{Tr}\,R_2(u_0)\ln R_2(u_0)
\end{equation}
provides the minimum value of information
\begin{equation}\label{1.14a}
I(u_0)=\Sigma(u_0)-S,
\end{equation}
where $S=-\mbox{Tr}\,\rho\ln\rho$.

If the matrix $\rho$ is the density matrix of a bipartite system $\rho(1,2)$,
and $R_1$ and $R_2$ are the density matrices of the first and second
subsystems, respectively, the quantum information is
\begin{equation}\label{1.15} I_q=-\mbox{Tr} \,R_1(u_{10})\ln
R_{1}(u_{10})-\mbox{Tr}\, R_2(u_{20})\ln R_{2}(u_{20})+\mbox{Tr}\,\rho\ln\rho,
\end{equation}
where $ R_{1}(u_{10})$ and $R_2(u_{20})$ are the diagonalized density matrices
$R_1$ and $R_2$, and $u_{10}$ and $u_{20}$ are local transforms such that
$u=u_{10}\times u_{20}$. Thus, the  difference of information
$\Sigma(u_0)-I_q$ provides a characteristic of the correlations related to
global and local transforms $u_0$ and $u_{10}\times u_{20}$. Analogous
characteristics can be introduced using deformed information and the deformed
subadditivity condition.

\section{Conclusions}
Concluding, we list our main results obtained in this paper.

We obtained new classical and quantum entropic inequalities for the systems
without subsystems. The new inequalities have the same form as known
inequalities for composite systems. For example, the Araki--Lieb inequality
provides the relation of the von Neumann entropy of the quantum state of a
bipartite system to the difference of the entropies of the subsystem states.
We found an analogous inequality for the single qudit state, e.g., we found
the inequality for the qutrit state; this inequality can be checked
experimentally.

We formulated new entropic inequalities for quantum system tomograms, which
use the properties of tomograms to be fair probability distributions. In view
of the facts mentioned above, the information and entropic relations, which
are known for classical probability distributions are also valid for quantum
system states described by the tomographic probability distributions,
including the $q$-entropic inequalities.

We obtained new $q$-entropic inequalities for tomograms of single qudit
states, e.g., for $j=3/2$. The physical meaning of the entropic inequalities
we obtained is related to the fact that they describe the properties of
quantum correlations in the single qudit state connected with quantum
fluctuations of the degrees of freedom like different spin projections in the
same system in contrast to bipartite systems where the inequalities are
related to the properties of quantum correlations of the degrees of freedom of
different subsystems.

The existence of quantum correlations in a single system formally is
completely analogous to the correlations in multipartite systems, that is
demonstrated by the existence of the entropic inequalities found here. The
inequalities can be used to elaborate the quantum resource for applications in
quantum technologies. Such a resource is usually considered as the resource of
quantum correlations in multipartite systems. In our study, we showed that a
resource is available in the systems without subsystems due to quantum
correlations of their degrees of freedom. This possibility was mentioned in
\cite{TurinMA}; it will be studied in a future publication.

\subsection*{Acknowledgements}
O.V.M. thanks the Organizers of the Conference "Quantum Theory: from Problems
to Advances" and especially Prof. A. Khrennikov for invitation and kind
hospitality.

\end{document}